\documentclass[reprint]{JASA}
\usepackage{subfigure}
\usepackage{algpseudocode}
\begin{document}
\title[]{Motion Acoustic Flow Field: Motion Estimation for Blob Targets in Active Sonar Echograph of Harbor Environments}
\author{Zhuoqun Wei}
\affiliation{School of Marine Science and Technology, Northwestern Polytechnical University, Xi'an 710072, China}
\author{Yina Han}
\email{Electronic mail: yina.han@nwpu.edu.cn, ORCID: 0000-0001-9713-1821.}
\affiliation{School of Marine Science and Technology, Northwestern Polytechnical University, Xi'an 710072, China}
\author{Shuang Zhao}
\affiliation{School of Marine Science and Technology, Northwestern Polytechnical University, Xi'an 710072, China}
\author{Qingyu Liu}
\affiliation{Naval Research Academy, Beijing 100161, China}
\author{Jun Song}
\affiliation{Naval Research Academy, Beijing 100161, China}
\date{\today}
\begin{abstract}
\indent Motion feature is of great significance for blob targets recognition, behavior analysis and threat estimation in active sonar echographs. Hence, it is desirable to access the space-time variation of echo intensity on each spatial-temporal resolution cell from the sonar echographs sequence. Then the subtle motion information of the potential blob targets can be accurately characterized. This idea has been conduced in optical image sequences by solving an motion optical flow field (MOFF) function. Nonetheless, due to the sparkle of the sonar echograph sequences, and strong interferences caused by wake and cavitation noise of fast-moving ship in harbor environments, the constraints underlying the traditional motion optical flow function that is couples the brightness constancy constant along time dimension of each echo intensity points and the motion field spatial smoothness does not hold in our case. Hence, this paper presents a new motion acoustic flow field (MAFF) function and its solving strategy to accurately characterize the subtle motion information of blob targets in active sonar echographs of harbor environments. Experiments on a series of cooperative targets in real-world harbor environments demonstrate the efficacy of our proposed MAFF.
\end{abstract}
\maketitle
\setlength{\belowcaptionskip}{0.1cm}
\section{\label{sec:1} Introduction}
\indent An important goal of sonar system design has been to track and recognize divers or unmanned underwater vehicles (UUVs) to ensure the security of vital military and commercial facilities \citep{53}. Given the challenge of hunting these quiet underwater intruders, it is more effective to obtain target echoes from active sonar with high frequency. Generally, sonar returns are normally displayed as an echograph, which reflects the intensity of the echoes in different ranges and bearings \citep{8}. Owing to the multipath effect, active sonar in harbor environments encounters strong reverberation arising from distributed (rough bottom structures) and discrete (bubbles, fish schools) scatters \citep{54,55}. High-level scattered returns in an echograph are often easily confused with actual target echoes \citep{56}. In addition, as details on the size and appearance signature of quiet underwater intruders are lacking, they exist as blob targets in active sonar echographs. For example, a potential invasion target (a small blob marked by a red rectangle on the active sonar echograph in Fig.~\ref{F1}(a)) may be confuse with other light blobs in such sonar echograph, where brighter blobs indicate higher intensity. Furthermore, diverse invasion targets (the small blob marked by a red rectangle on the active sonar echograph in Fig.~\ref{F1}(a) and Fig.~\ref{F1}(b)) have similar size and appearance signature (light blobs) in sonar echograph, which makes it difficult to effectively distinguish from each other. Thus, there is a need for effective methods to characterize targets in active sonar echographs. \\
\indent Different from imaging sonar \citep{9}, the motion feature of blob targets can be effectively characterize targets in active sonar echographs. For example, target tracking is commonly used to refine motion feature into a more useable form at the track level, producing estimates of the likely number of targets and their position, course, and speed. Estimates of target characteristics, such as target type, may also be produced \citep{61}. In addition, If motion feature is available with the other target characteristics and the correlation between a feature value and the target classification is strong, the feature can be used to reject target before tracking \citep{62}. SNR and Doppler are commonly used in this role \citep{63}. However, Due to the influence of high level littoral clutters, the direct extraction of blob targets motion information by using the tracker will fail. While the most effective method of motion feature extraction so far is Doppler estimation \citep{8,30}, but in order to overcome the clutter interference, these method needs to carry on the fine design of the active sonar transmitting signal. In our previous work, a high-order time lacunarity (HOT-Lac) feature is proposed for detecting potential targets from high level littoral clutters in active sonar echographs \citep{4}, which can obtain motion information of blob targets, but it involves high computational complexity of loop summations. Although the algorithm has been improved in \citep{60} in later studies, but the new algorithm still has to consume a period of time to accumulate enough frames of sonar echographs for the start of calculation. Hence, motion feature is of great significance for blob targets recognition, behavior analysis and threat estimation in active sonar echographs. It is desirable to access the space-time variation of echo intensity on each spatial-temporal resolution cell from the sonar echographs sequence. Then the subtle motion information of the potential blob targets can be accurately characterized. \\
\indent This idea has been conduced in optical image sequences by solving an motion optical flow field (MOFF) function in the field of computer vision \citep{41,17}. As a typical optical flow approach, HS method assumed that the brightness of a particular point in the pattern is constant between image sequence and the apparent velocity of the brightness pattern varies smoothly almost everywhere in the image. HS method coupled the brightness constancy and MOFF spatial smoothness assumptions using an energy function called MOFF function, which included data term and spatial term formulated by quadratic both. Based on the HS method, researchers have carried out fruitful researches on solving specific problems of MOFF function, which were fall into two branches. The first branch is to improve the optical flow function in order to enhance its robustness. The early improvement of MOFF function is mainly aimed at the improvement of penalty function. Due to the quadratic penalty function is not robust to outliers, \citet{18} used an L1 penalty function instead to preserve flow discontinuities. \citet{19} used a robust penalty function, Lorentzian, which is a non-convex function. \citet{22} and \citet{20} used the Charbonnier, the most robust convex function. The widely used Charbonnier function to penalty a differentiable variant of the absolute value. While the recent work is mainly by improving the local spatial smoothing term. \citet{22, 28} added a non-local smoothness term to MOFF function, that robustly integrates flow estimates over large spatial neighborhoods. The second branch is to propose efficient and accurate computing frameworks. In terms of efficiency, the current mainstream computing framework is a coarse-to-fine, warping-based approach which is called image pyramid \citep{20, 22}. \citet{24} theoretically justified the warping-based estimation process. As it is computationally impractical to perform a full search of a image in previous motion optical flow computing frameworks, image pyramid saves computing resources dramatically. In terms of accuracy, one idea is to use $Rudin - Osher - Fatemi$ (ROF) structure texture decomposition method \citet{21} and their linear combinations \citet{28, 22, 35}. This method effectively resists the influence of brightness changes on the estimation accuracy of motion optical flow. Simpler alternatives, the experimental results show that filter response (or high-order) constancy \citep{24, 26, 51, 36} can serve the same purpose as ROF texture decomposition. Therefore, the current research does not have a clear conclusion as to which image pre-processing method can achieve better results in different scenarios. \\
\indent Inspired by the idea of optical flow, we attempted to propose a similar motion field estimation approach to estimate the motion feature of small targets in harbor environments. Nonetheless, due to the sparkle of the sonar echograph sequences, and strong interferences caused by wake and cavitation noise of fast-moving ship in harbor environments, the constraints underlying the traditional motion MOFF function that is couples the brightness constancy constant along time dimension of each echo intensity points and the motion field spatial smoothness does not hold in our case. Thus, the existing optical flow method can not be applied directly in harbor environments. Hence, this paper presents a new motion acoustic flow field (MAFF) function and its solving strategy to accurately characterize the subtle motion information of blob targets in active sonar echographs of harbor environments. The MAFF improves non-local term by adding regularization parameters based on context information, and the pre-processing method is evaluated comprehensively. Furthermore, experiment results on sets of real-world datasets from the South China Sea demonstrate the potential of MAFF in small targets motion feature estimation. \\
\indent The remainder of the paper is organized as follows. Section~\ref{sec:2} presents the MAFF estimation method. Section~\ref{sec:3} describes the sonar data set used in the experiments. Section~\ref{sec:4} presents the evaluation metrics, the discussion of the implementation details in MAFF estimation, and performance on small targets motion feature estimation. Our conclusions are presented in Sec.~\ref{sec:5}.
\section{\label{sec:2} Motion acoustic flow field}
\subsection{\label{subsec:2:1} Motion optical flow field}
\subsubsection{\label{subsec:2:1:1} Computation framework of MOFF}
\begin{figure*}
\begin{center}
\includegraphics[width = .9\textwidth]{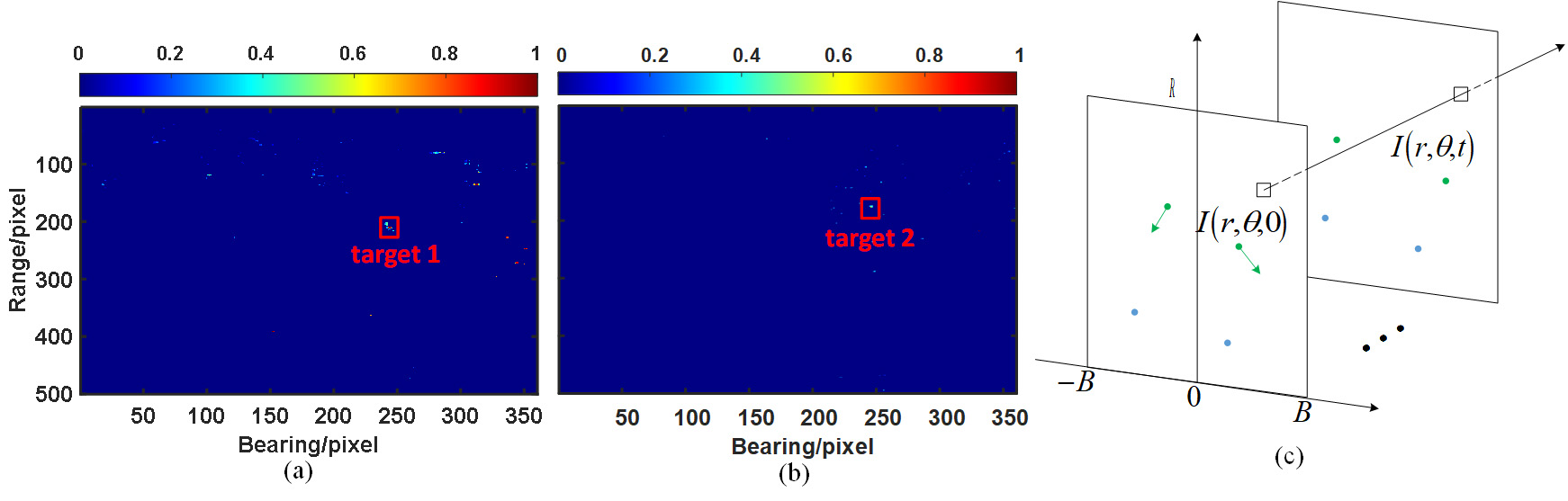}
\caption{\label{F1}{Active sonar echographs in real-world harbor environments and outline of active sonar echograph sequences: (a) active sonar echograph with potential invasion blob target 1; (b) active sonar echograph with potential invasion blob target 2. Blob target is marked by a red rectangle on the active sonar echograph. (c) outline of active sonar echograph sequence, green spots denote moving blobs and blue spots denote static blobs. }}
\end{center}
\end{figure*}
\begin{figure*}
\begin{center}
\includegraphics[width = .9\textwidth]{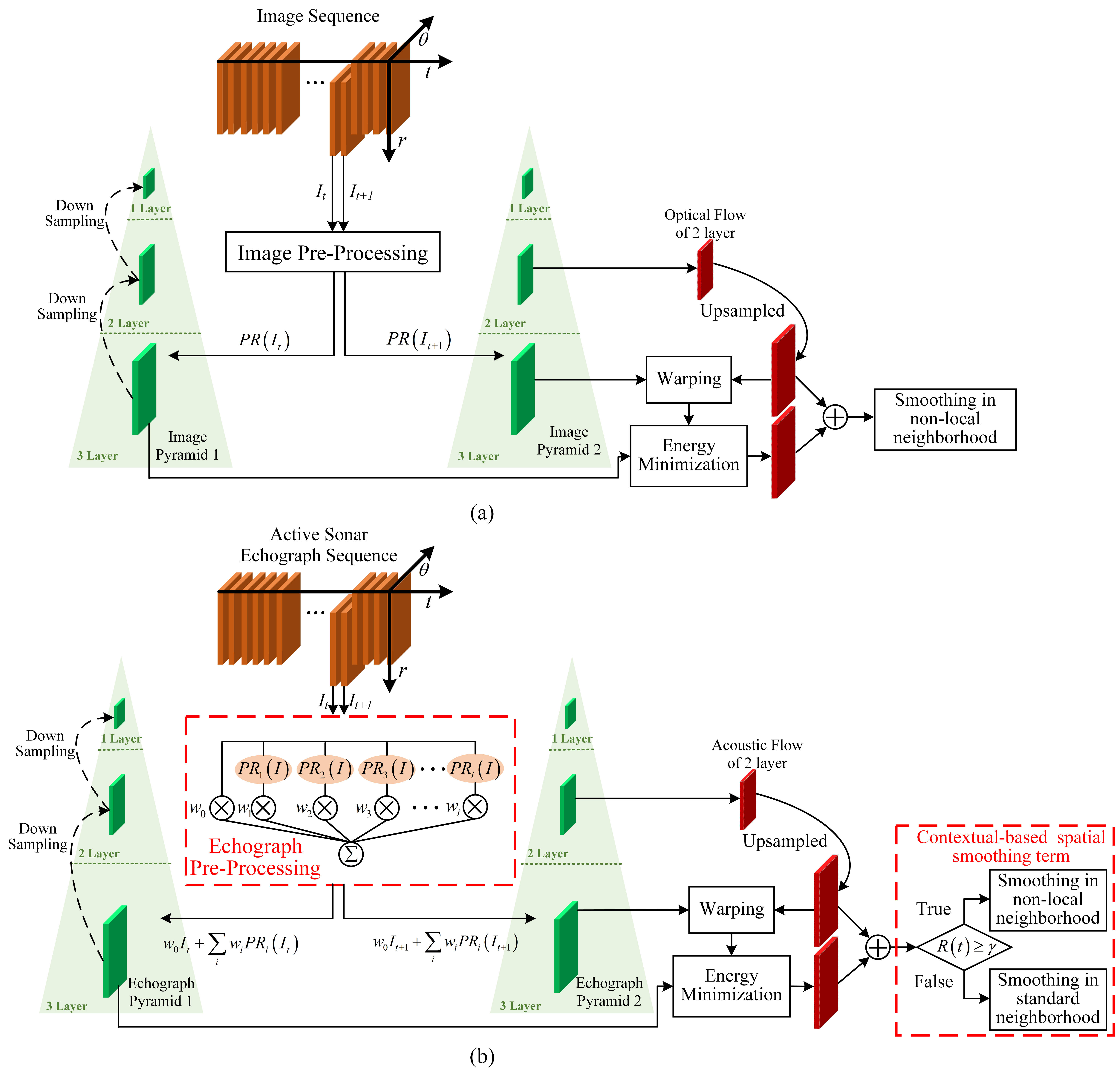}
\caption{\label{F2}{Computation framework of MOFF and MAFF: (a) Computation framework of MOFF includes fixed image pre-processing and image pyramid processing which MOFF estimated in each pyramid layer is smoothed in non-local neighborhood; (b) Computation framework of MAFF includeds combination pre-processing method for echographs and echograph pyramid processing which MAFF estimated in each pyramid layer is smoothed with contextual-based spatial smoothing term.}}
\end{center}
\end{figure*}
\indent Figure~\ref{F2}(a) shows the coarse-to-fine, warping-based computation framework of MOFF. As shown in Fig.~\ref{F2}(a), in each MOFF calculation, image pre-processing is used on two adjacent frames $I_{t}$ and $I_{t+1}$ in the image sequence, and the pre-processed images $PR(I_{t})$ and $PR(I_{t+1})$ are used to construct image pyramid 1 and image pyramid 2 respectively (the green triangle in Figure~\ref{F2} shows the example of a three-layer image pyramid). Each layer of the image pyramid is recursively down-sampled from its nearest lower layer. The resolution of each layer in the image pyramid decreases with the decrease of the layer's number. MOFF computed at a coarse level is used to warp the second image toward the first at the next finer level. The MOFF is obtained from the top of the pyramid layer by layer, that is, the coarse-to-fine, warping-based computation framework. The operations performed at the bottom of the pyramid are shown on the right of Figure~\ref{F2}(a), which are the same as the calculations in each of the remaining pyramid layers.
\subsubsection{\label{subsec:2:1:2} Motion optical flow function}
\indent As shown in \citet{22}, the motion MOFF function can be defined as
\begin{equation}
\begin{aligned}
   E(\mathbf{u},\mathbf{v})={}&\sum_{x=1}^{X}\sum_{y=1}^{Y}  \rho_D \big(PR(I_{t}(x,y))\\
   &-PR(I_{t+1}(x+u_{x,y},y+v_{x,y})) \big) \\
   &+\sum_{x=1}^{X}\sum_{y=1}^{Y}  \lambda_{DS} \bigl[ \rho_S (u(x,y)-u(x+1,y))\\ 
   &+ \rho_S (u(x,y)-u(x,y+1))  \\
   &+\rho_S (v(x,y)-v(x+1,y))\\ 
   &+ \rho_S (v(x,y)-v(x,y+1)) \bigr]   \\
   &+\lambda_{N}\sum_{x,y}\sum_{(x^{'},y^{'}) \in \mathcal{N}_{x,y}}  \left(\left|u(x,y)-u(x^{'},y^{'})\right| \right. \\
   &+\left.\left|v(x,y)-v(x^{'},y^{'})\right|\right),
\end{aligned}
\label{E1}
\end{equation}
where $I_{t}$ is a $X \times Y $ image, $\mathbf{u}$ and $\mathbf{v}$ are the horizontal and vertical components of the MOFF, $\mathbf{u}=\{u(x,y)|1\leq x\leq X,~1\leq y\leq Y\}$, $\mathbf{v}=\{v(x,y)|1\leq x\leq X,~1\leq y\leq Y\}$. The first term in Eq.~\ref{E1} is data term, which assumes brightness constancy, $PR(I_{t})$ and $PR(I_{t+1})$ are two image frames, $(x,y)$ indexes a particular image pixel location. The second term in Eq.~\ref{E1} is spatial term, which assumes motion optical flow field spatial smoothness, $u_{x,y}$ and $v_{x,y}$ are elements of $\mathbf{u}$ and $\mathbf{v}$ respectively. $\rho_D$ is data penalty function, $\rho_S$ is spatial penalty function
\begin{equation}
\begin{aligned}
   & \rho_D(I_{s})=\left(I_{s}^{2}+\epsilon^{2}\right)^{a}, \\
   & \rho_S(I_{s})=\left(I_{s}^{2}+\epsilon^{2}\right)^{a},
\end{aligned}
\label{E2}
\end{equation}
where, $\epsilon=10^{-3}$, and $a=0.45$, which are the optimal parameter investigated in~\citep{22}. $\lambda_{DS}$ is a regularization parameter that balance brightness constancy and motion field spatial smoothness. The last term in Eq.~\ref{E1} is a non-local term, which imposes a particular smoothness assumption within a specified large region of the motion optical flow field, $\mathcal{N}_{x,y}$ is the set of neighbors of pixel $(x,y)$ in a possibly large area and $\lambda_{N}$ is a scalar weight. The non-local term minimizes the L1 distance between the central value and all motion optical flow values in its neighborhood except itself.
\subsection{\label{subsec:2:2} Motion acoustic flow field}
\subsubsection{\label{subsec:2:2:1} Computation framework of MAFF }
\begin{figure}
\begin{center}
\includegraphics[width = .25\textwidth]{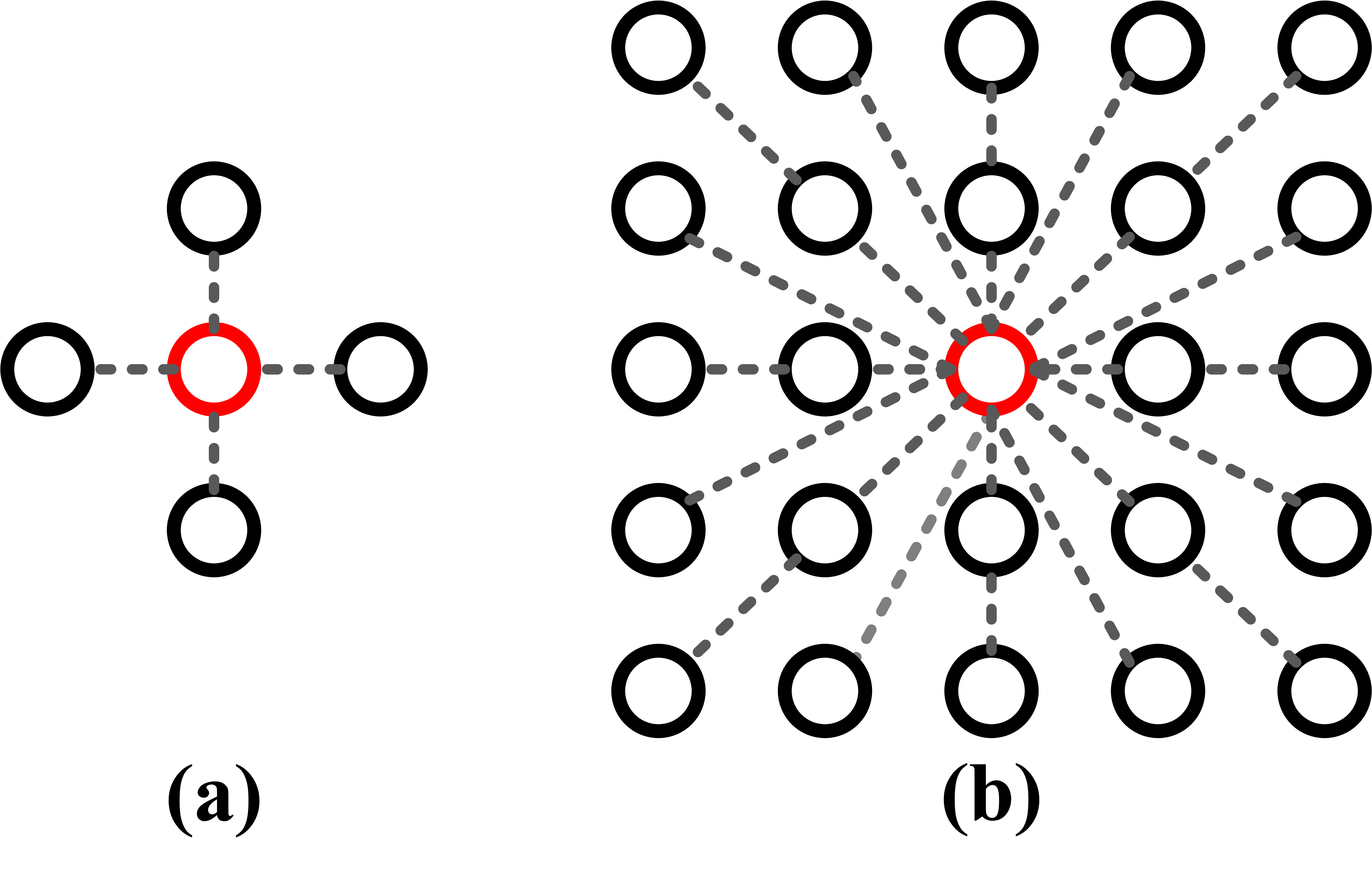}
\caption{Neighborhood structure for the center (red) pixel: (a) the standard pairwise neighborhood model; (b) the non-local neighborhood model}
\label{F3}
\end{center}
\end{figure}

\indent Figure~\ref{F2}(b) shows the computation framework of MAFF proposed in this paper. Different from the typical MOFF computational framework given in Figure~\ref{F2} (a), the proposed novel MAFF computation framework optimizes pre-processing module and the spatial smoothing module to hold our case (the red dotted box in Figure~\ref{F2} (b) shows our method). Firstly, in the echograph pre-processing module, MAFF computation framework adopted the combination pre-processing method, which linearly combined the results of the original echograph and several different pre-processing methods to obtain the final pre-processing results. Specifically, several different pre-processing methods is ranging from ROF structure texture decomposition, Gaussian pre-filter, Laplacian pre-filter, LoG pre-Filter and Sobel edge detector, these methods have been proved to be effective in the previous work and our previous experiments. In addition, the weight $w_{I}$ of the linear combination is determined by experiment, which will be expanded in detail in \ref{subsubsec:4:2:2}. Secondly, in the spatial smoothing module, we introduce global context information to determine whether to perform spatial smoothing for MAFF in the non-local neighborhood or in the standard neighborhood. Figure~\ref{F3} shows the difference between the two neighborhoods, the standard pairwise neighborhood model connects a center pixel with its nearest neighbors, while the non-local term connects a pixel with many pixels in a large spatial neighborhood. The basis for the judgment is formulated in the next section.
\subsubsection{\label{subsec:2:2:2} Motion acoustic flow function}
\indent The intensity of the target echo is generally a function of the size of the entity and the severity of the impedance changes, in most cases providing an easily quantifiable measure of its range and bearing from a receiver. The correlativity between incident acoustic intensity and target echo intensity of active sonar is
\begin{equation}
\frac{I_s}{I_i} \propto L^2\left(\dfrac{\sin{\beta}}{\beta}\right)^2 \cos^2{\alpha}
\label{E3}
\end{equation}
where $I_{s}$ is the target echo intensity, $I_{i}$ is the incident acoustic intensity. $L$ is the length of the longer side of the scatterer, $\beta \propto L\sin{\alpha}$, where $\alpha$ is the angle between the incident direction of the acoustic wave and the normal direction. As shown in Fig.~\ref{F1}(c), HF active sonar echo signals are displayed as echograph
\begin{equation}
\begin{aligned}
I_{t}(r,\theta) = I_{s}(r,\theta,t),
\end{aligned}
\label{E4}
\end{equation}
where $I_{t}$ is active sonar echograph at time t, $(r,\theta,t)$ represents a certain spatial-temporal resolution cell of sonar. And at the time $t$, $(r,\theta)$ is a certain pixel in sonar echograph. As shown in Figure~\ref{F2}(b), pre-processed sonar echograph is
\begin{equation}
\begin{aligned}
P_{t}(r,\theta) = w_0I_t+ \sum_{i}w_iPR_i(I_{t}(r,\theta)),
\end{aligned}
\label{E5}
\end{equation}
where $w_0$ is weight of original echograph, $w_i (i=1,2,3...)$ are weights of pre-processing methods, $PR_i(\dot) (i=1,2,3...)$ are results of pre-processing methods. The motion acoustic flow function defined as
\begin{equation}
\begin{aligned}
   E(\mathbf{u},\mathbf{v})={}&\sum_{r=0}^{R}\sum_{\theta=-B}^{B}  \rho_D \big(P_{t}(r,\theta)-P_{t+1}(r,\theta) \big) \\
   &+\sum_{r=0}^{R}\sum_{\theta=-B}^{B}  \lambda_{DS} \bigl[ \rho_S (u(r,\theta)-u(r+1,\theta))\\
   &+\rho_S (u(r,\theta)-u(r,\theta+1))  \\
   &+\rho_S (v(r,\theta)-v(r+1,\theta))  \\
   &+\rho_S (v(r,\theta)-v(r,\theta+1)) \bigr] \\
   &+\lambda_{N}{'}\sum_{r,\theta}\sum_{(r^{'},\theta^{'}) \in \mathcal{N}_{r,\theta}}  \left(\left|u(r,\theta)-u(r^{'},\theta^{'})\right|\right.\\
   &+\left.\left|v(r,\theta)-v(r^{'},\theta^{'})\right|\right),
\end{aligned}
\label{E6}
\end{equation}
where $R$ and $B$ are the beam and range limits. $\mathcal{N}_{r,\theta}$ is neighborhood of $(r,\theta)$ in the sonar echograph, $(r^{'},\theta^{'})$ are pixels in $\mathcal{N}_{r,\theta}$. In Eq.~\ref{E6}, an adaptive regularized parameter $\lambda_{N}{'}$ is used to replace the original fixed regularized parameter [$\lambda_{N}$ in Eq.~\ref{E1}]. At the time $t$,
\begin{equation}
\lambda_{N}{'}=\left\{
\begin{aligned}
& 1,&\ & if&\ R(t) \geq \gamma \\
& 0,&\ & if&\ R(t) < \gamma,
\end{aligned}
\right.
\label{E7}
\end{equation}
where, the threshold $\gamma$ is the threshold value. As the cavitation noise appear as vertical strips in sonar echograph, the appearance of fast-moving ships intrude into the detection area will result in a large number of false target points. And $\gamma$ can be determined by statistical analysis of false target points in the same scenario in the previous experimental data. $R(t)$ is the ratio of foreground pixels of all pixels
\begin{equation}
R(t) = \frac{1}{2B \times R}\sum_{r=-B}^{B}\sum_{\theta =0}^{R}I_{b}(r,\theta,t).
\label{E8}
\end{equation}
\indent To obtain global context information, the sonar echograph is characterized by a binary image whose pixel values are given by
\begin{equation}
I_{b}(r,\theta,t)=\left\{
\begin{aligned}
& 1,&\ & if&\ (r,\theta,t) \in P_{F} \\
& 0,&\ & if&\ (r,\theta,t) \in P_{B},
\end{aligned}
\right.
\label{E9}
\end{equation}
where $P_{F}$ and $P_{B}$ are collections of foreground pixels (potential moving pixels) and background pixels (static pixels) respectively which are split by a normal Gaussian mixture model. In this paper, the second and third terms in Eq.~\ref{E6} are collectively referred to as a spatial smooth term in the motion acoustic flow function.
\section{\label{sec:3} Sonar data set}
\indent \indent In our experiment, we use the multi-beam active sonar system with a fixed shore-based platform for harbor monitoring. The periodic acoustic pulse is transmitted into the underwater environment and moving targets and then collected by receiving arrays. An active sonar echograph is a 2D grayscale image composed of received acoustic signals, which are beamformed, match filtered, and normalized. Here, using normalized sonar data after gating tests, we discuss the ability of MAFF to characterize the subtle motion information of blob targets in active sonar echographs of harbor environments.
\begin{figure}
\begin{center}
\includegraphics[width = .48\textwidth]{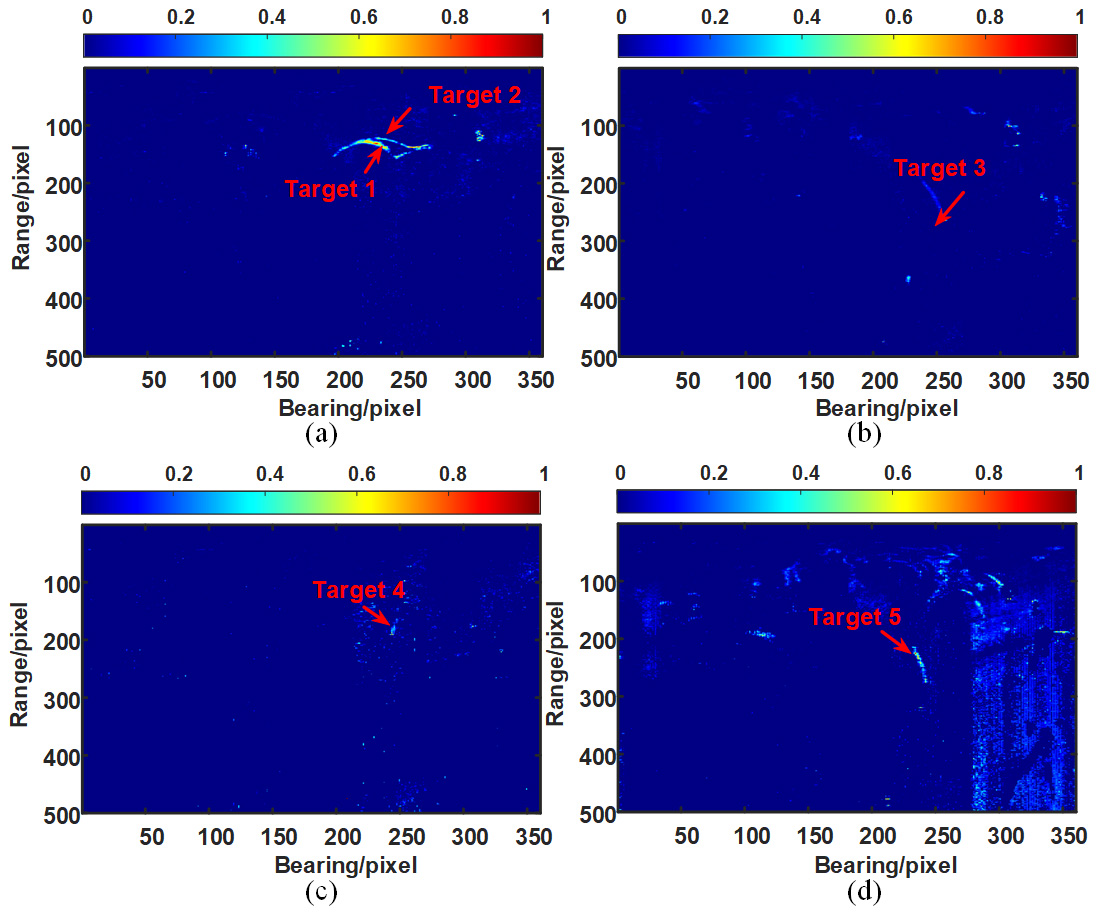}
\caption{Normalized multiframe accumulation from active sonar echographs, in which the targets of interest are moving in the red arrows: (a) Data set 1, including 130 frames, Small target 1 towed Small target 2 in a curved motion across the active sonar beam, the echo energy of the Small target 1 was higher than that of the Small target 2; (b) Data set 2, including 80 frames, Small target 3 were moving in an approximately straight line away from the sonar; (c) Data set 3, including 40 frames, Small target 4 was moving in an approximately straight line away from the sonar, with this motion being slower than that of the targets in Data set 2; (d) Data set 4, including 188 frames, Small target 5 also moved in an approximately straight line away from the sonar. }
\label{F4}
\end{center}
\end{figure}
\indent Generally, the performance evaluation data can be obtained through numerical modeling and real experiments. However, both analytical and numerical modeling are difficult because reverberation in real oceans is highly non-Gaussian \citep{30}. Therefore, real experimental data is used for performance evaluation in this paper. The echographs used in this paper were drawn from real measured sonar data collected from a single region by multiple experiments in the South China Sea in 2015. Specifically, four representative sets of images were collected, containing different moving targets that are of concern in a harbor environment. The targets were chosen to have different echo energy, object size, and motion mode in order to verify the robustness of the proposed feature. These active sonar echographs are shown in Fig.~\ref{F4} (multi-frame accumulation of active sonar echographs can be obtained by accumulating the echo on each spatial-temporal resolution cell, the echographs are then normalized by echo intensity).
\section{\label{sec:4} Experimental evaluation}
\subsection{\label{subsec:4:1} Evaluation metrics}
\indent The aim of this paper was to evaluate whether the MAFF could accurately characterize the subtle motion information of blob targets in active sonar echographs of harbor environments. In consideration of the practical needs of motion information observation, we explicitly show the motion information by coding the MAFF with a specific color key. For example, Fig.~\ref{F5} (a) shows vector $(u,v)$ in MAFF, consider the polar coordinates case:
\begin{figure}
\begin{center}
\includegraphics[width = .45\textwidth]{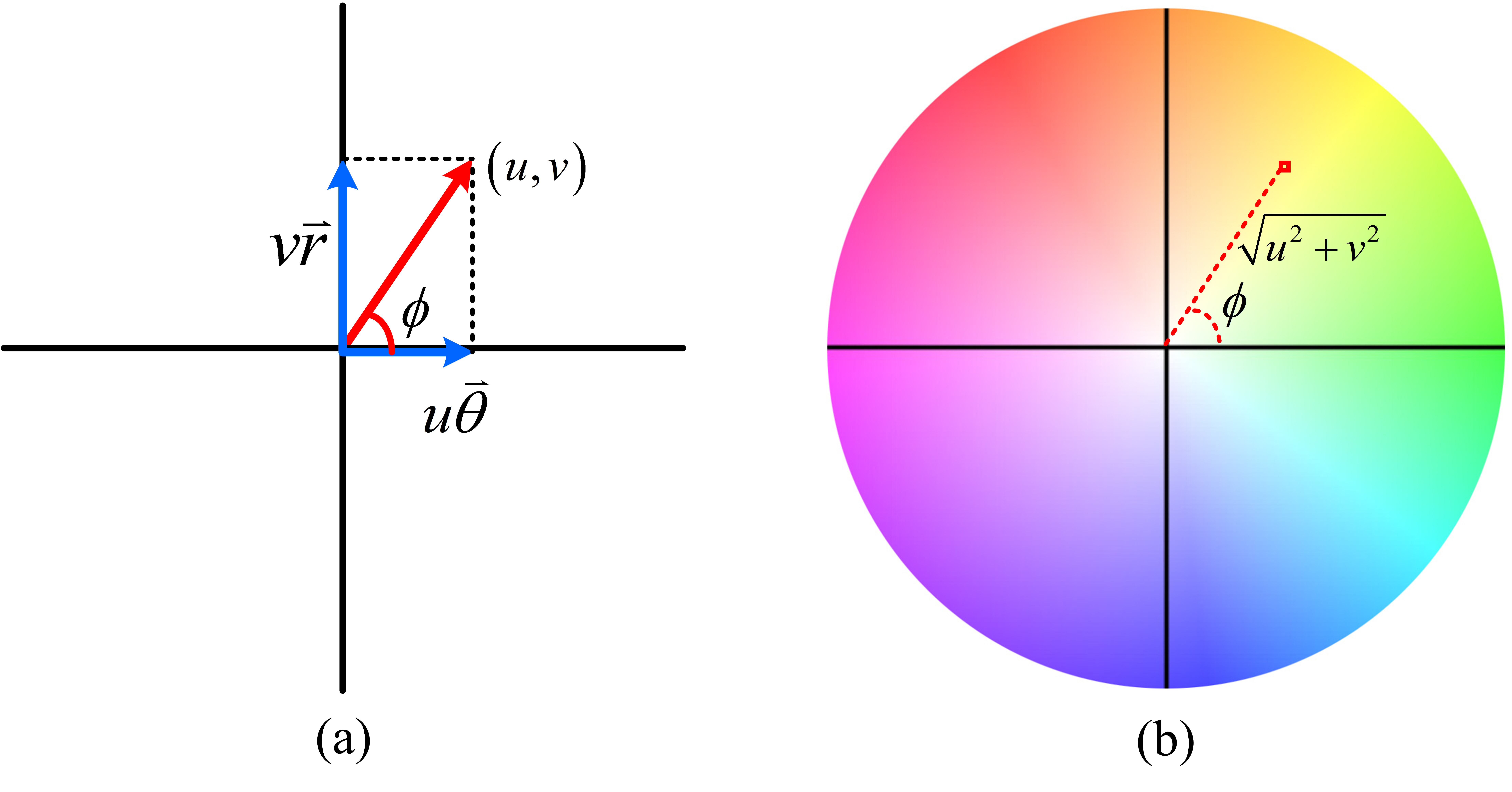}
\caption{\label{F5}{(a) MAFF in cartesian coordinates; (b) MAFF in polar coordinates, define a color key of MAFF, $(\mathbf{u},\mathbf{v})$ has a unique mapping in the color key map, that is, to distinguish the different direction of motion by hue, to distinguish the speed of motion by color saturation.}}
\end{center}
\end{figure}
\begin{figure*}
\begin{center}
\includegraphics[width = .8\textwidth]{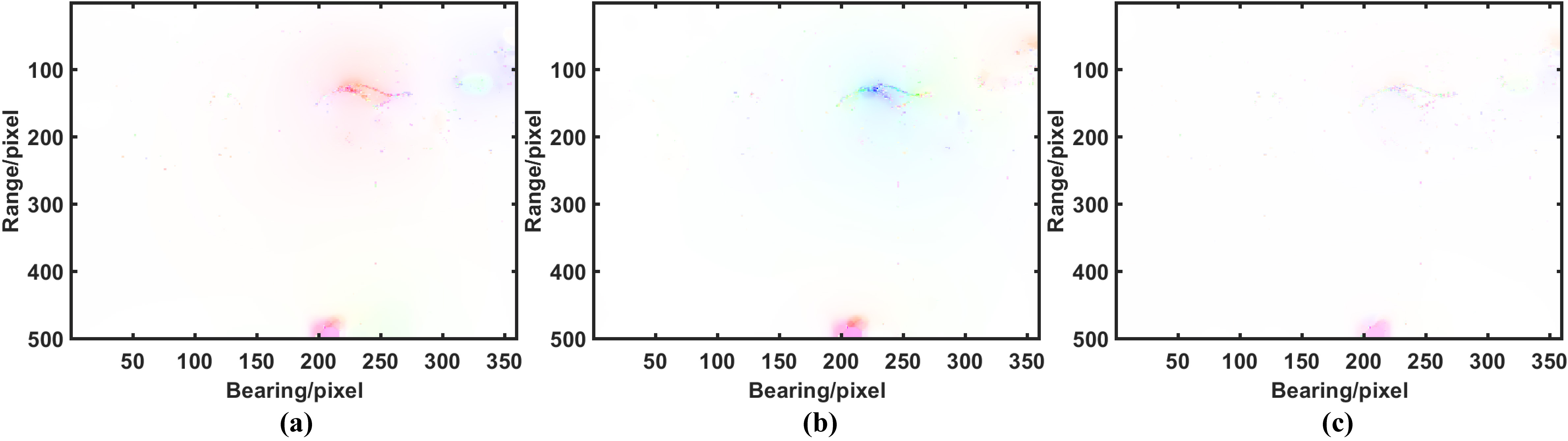}
\caption{\label{F6}{MAFF map (normalized multiframe accumulation of coded MAFF): (a) the forward MAFF map, the displacements from the pixels in frame $t$ to theirs corresponding pixels in the frame $t+1$; (b) the backward MAFF map, the displacements from the pixels in frame $t+1$ to theirs corresponding pixels in the frame $t$ ; (c) the forward-backward error (FBE), the sum of forward MAFF and backward MAFF, that an ideal correspondence should be zero. }}
\end{center}
\end{figure*}
\begin{equation}
\begin{aligned}
|(u,v)|&= \sqrt{u^2+v^2} \\
\phi ~~~~&= \arctan \frac{v}{u} ,\ 0\leq\phi\leq2\pi,
\end{aligned}
\label{E10}
\end{equation}
where $|(u,v)|$ is the speed of blob target moving on active sonar echograph, $\phi$ is direction of motion. As shown in Fig.~\ref{F5}(b), the magnitude and direction of flow can be uniquely mapped to color saturation and hue in the color key map. Nevertheless, subtle visual differences usually affect the judgment of the motion information of the blob targets. Hence, the industry usually uses the motion flow field estimation error for quantitative assessment. Generally, flow field estimation error can be measured in two main ways. One is to obtain average angular error (AAE) and average end-point error (EPE) by comparing estimated motion flow field with ground truth motion flow field. Another is the forward-backward error (FBE) \citep{32, 33}, which is a way to compute flow field estimation error without ground truth motion flow field. In our experiment, it is almost impossible to obtain the ground truth motion flow field, because it is difficult to manually label the correct pixel wise correspondences of active sonar echographs measured in real experiments data even if the prior information such as speed and direction of targets have been known. Therefore, we use average forward-backward error (AFBE) for quantitative assessment. AFBE of frame $t$ in echograph sequence is
\begin{equation}
\begin{aligned}
\text{FBE}_{t} = {}&\frac{1}{2B \times R}\sum_{r=-B}^{B}\sum_{\theta=0}^{R} \left| \mathop{u} \limits ^{\rightarrow} (r,\theta)+\mathop{u}\limits ^{\leftarrow} (r,\theta)\right.\\
&+\left.\mathop{v}\limits ^{\rightarrow} (r,\theta)+\mathop{v}\limits ^{\leftarrow} (r,\theta) \right|,
\label{E11}
\end{aligned}
\end{equation}
where, $(\mathop{u}\limits ^{\rightarrow}(r,\theta),\mathop{v}\limits ^{\rightarrow}(r,\theta))$ is estimated by the pixel $(r,\theta)$ in the frame $t$ to its corresponding pixel in the frame $t+1$, and $(\mathop{u}\limits ^{\leftarrow}(r,\theta),\mathop{v}\limits ^{\leftarrow}(r,\theta))$ is estimated by the pixel $(r,\theta)$ in the frame $t+1$ to its corresponding pixel in the frame $t$. As shown in Figs.~\ref{F6}(a) and (b), the forward MAFF $(\mathop{\mathbf{u}}\limits ^{\rightarrow},\mathop{\mathbf{v}}\limits ^{\rightarrow})$ ( $\mathop{\mathbf{u}}\limits ^{\rightarrow}=\{\mathop{u} \limits ^{\rightarrow} (r,\theta)|0\leq r\leq R, -B\leq \theta\leq B\}, \mathop{\mathbf{v}}\limits ^{\rightarrow}=\{\mathop{v} \limits ^{\rightarrow} (r,\theta)|0\leq r\leq R, -B\leq \theta\leq B\}$) and the backward MAFF $(\mathop{\mathbf{u}}\limits ^{\leftarrow},\mathop{\mathbf{v}}\limits ^{\leftarrow})$ ($\mathop{\mathbf{u}}\limits ^{\leftarrow}=\{\mathop{u} \limits ^{\leftarrow} (r,\theta)|0\leq r\leq R, -B\leq \theta\leq B\}, \mathop{\mathbf{v}}\limits ^{\leftarrow}=\{\mathop{v} \limits ^{\leftarrow} (r,\theta)|0\leq r\leq R, -B\leq \theta\leq B\}$) are reversed. The forward-backward error is the sum of forward MAFF and backward MAFF, which is zero theoretically [Fig.~\ref{F6}(c)]. It's worth noting that we use normalized multiframe accumulation of coded MAFF to construct MAFF map to display the motion information of the blob target in the multi-frame sonar echograph in a single graph.
\subsection{\label{subsec:4:2} Ablation experiments: discussion of the implementation details in MAFF estimation}
\indent The main innovations of this paper are combination pre-processing method and contextal based MAFF spatial smoothing method. In order to analyze the enhanced motion estimation accuracy achieved by our work, in this section
we introduce the concept of ablation analysis, a procedure investigation that is conducted by starting from a basic configuration then iteratively modifying one setting of the model and seeing how that affects performances on the sonar datasets. Specifically, in Sec.~\ref{subsubsec:4:2:1}, we discuss the ability of combination pre-processing method with spatial smoothing method based on non-local term, while in Sec.~\ref{subsubsec:4:2:2}, we discuss the ability of contextal based MAFF spatial smoothing method with combination pre-processing method. The real experimental data and evaluation metrics introduced in Sec.~\ref{sec:3} and Sec.~\ref{subsec:4:1} will be used in this section.
\subsubsection{\label{subsubsec:4:2:1} Pre-processing method}
\indent In this part, the pre-processing method in estimation framework of MAFF from the perspective of the accuracy of MAFF estimation are discussed. To gain accurate estimation of optical flow field against brightness changes ROF structure texture decomposition method to reduce the influence of illumination changes between frames. As the improvement linearly combined the texture and structure components (in the proportion 20:1). Simpler alternatives, a variety of pre-filters had been used in previous research, for example, Laplacians pre-filter boosted the high frequency while suppressing the low frequency components that contain the lighting change. Nevertheless, Laplacians pre-filter is sensitive to noise. Gaussians pre-filter has good performance in denoising, but it could easily over-smooth image sequences. Meanwhile, Gaussians pre-filter performed well on processing the synthetic image sequences, but poorly on real ones . The LoG filter combined the advantages of these two, and could be generated by taking the Laplacian of the Gaussian kernel. Furthermore, edges had also been emphasized using the Sobel edge magnitude , but appeared to not work well on some image sequences, particularly the synthetic ones \citep{22}. \\
\indent Specifically, MAFF estimation approach without pre-processing is the basic configuration in this part, which is defined as a baseline model 1. Based on Baseline Model 1, the pre-processing method is iteratively modified, including ROF structure texture decomposing, Gaussian pre-filtering, Laplacian pre-filtering, LoG pre-filtering and Sobel edge detecting. For each modification, the estimation accuracy that evaluated by AFBF for each group of targets is shown in Table~\ref{T2}. The Bold entries highlight the optimal pre-processing method. Comparing the average AFBF for each modification, it can be seen that the LoG pre-filtering method provides the best performance of MAFF estimation. This can be attributed that LoG is robust to Gaussian noise while highlighting the edges of the target. \\
\begin{table*}
\caption{\label{T2}AFBFs of different pre-processing methods.}
\begin{ruledtabular}
\begin{tabular}{lccccc}
&\multicolumn{5}{c}{Data set} \\
\cline{2-6}
                                     &Data 1 & Data 2 &Data 3 & Data 4 & Average\\
\hline
No pre-process                       &0.0169 & 0.0154 &0.0227 & 0.1288 & 0.04595\\
\hline
ROF structure texture decomposition  &$\mathbf{0.0052}$ & 0.0061 &0.0071 & 0.0155 & 0.0085\\
Gaussian pre-filter                  &0.0128 & 0.0139 &0.0134 & 0.0413 & 0.02035\\
Laplacian pre-filter                 &0.0134 & 0.0150 &0.0138 & 0.0191 & 0.01533\\
LoG pre-filter                       &0.0062 & $\mathbf{0.0053}$ & $\mathbf{0.0070}$ & $\mathbf{0.0150}$ & $\mathbf{0.00838}$\\
Sobel edge detector                  &0.0376 & 0.0332 &0.0594 & 0.1390 & 0.0673 \\
\end{tabular}
\end{ruledtabular}
\end{table*}
\indent However, for each group of targets, owing to the different motion modes, there exists an optimal pre-processing method. Specifically, the ROF structure texture decomposition provides the lowest value of AFBF for Data set 1, which has more curves in the trajectory. The minimum value of AFBF for Data sets 2 to 4 exists on the pre-processing method of LoG pre-filter. As shown in Table \ref{T2}, the sobel edge detector achieves highest AFBFs whether in each individual data set or the average, which are significantly worse than the baseline approach. And Gaussian pre-filter and Laplacian pre-filter gains the similar performance, that neither of these methods differ significantly from the approach without pre-process. These results suggest that although constant target edge is very important for subsequent processing, but the pre-processing method that simply detects the edge without considering the influence of noise will often get worse results than that without pre-processing.
\subsubsection{\label{subsubsec:4:2:2} Spatial smoothing term}
\begin{figure}
\begin{center}
\includegraphics[width = .45\textwidth]{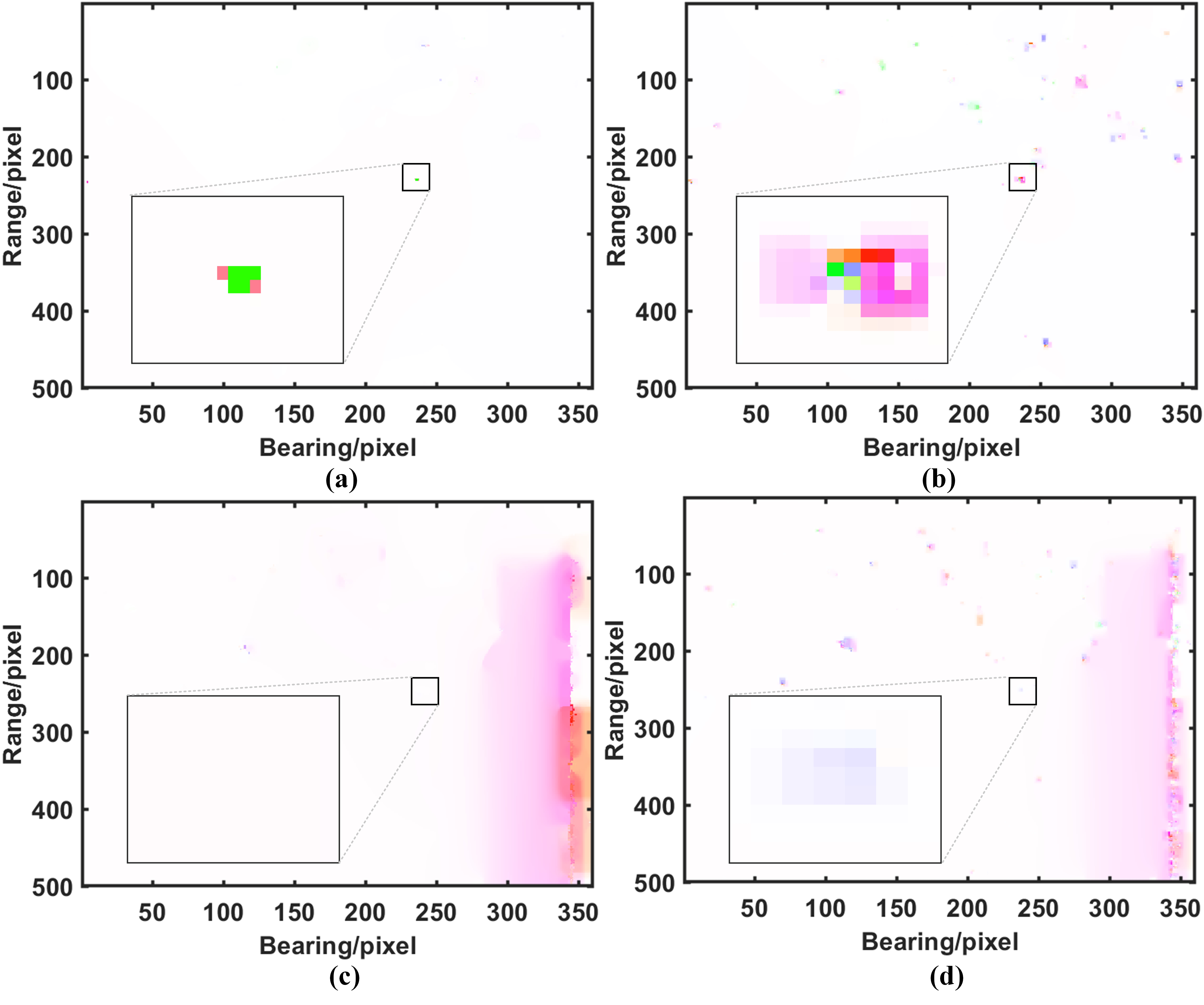}
\caption{\label{F7}{MAFF of small target 5: (a) w/o non-local term \& w/o fast-moving ship; (b) With non-local term \& w/o fast-moving ship;
(c) w/o non-local term \& with fast-moving ship; (d) With non-local term \& with fast-moving ship. Magnified versions of the black box given in the lower
left corner of the image. }}
\end{center}
\end{figure}

\indent In this part, we will provide some insight into how the MAFF estimation accuracy varies as the spatial smoothing term changes. Owing to the fast-moving ships that are not of interest intrude into the detection area, additional motion acoustic flow caused by the ship and its following stern wake was introduced artificially into the motion estimation of a interested target. As the wake and cavitation noise of fast-moving ship coexist with the target of interest in sonar echograph, the assumption of velocity of the brightness pattern varies smoothly within a large neighborhood will cause loss of target motion feature. The motion optical flow of small target has no absolute dominant position in the high level littoral background, which is easy to assimilate. \\
\indent As shown in Table \ref{T2}, the AFBFs of Data set 4 are significantly higher than others. To provide further detail, we explicitly show the MAFF of small target 5 that include the frame with additional fast-moving ship and without additional fast-moving ship, as shown in Fig~\ref{F7}. The target in Fig~\ref{F7}(a) is "clearer" than that in Fig~\ref{F7}(b). It is because of the role of non-local term in the motion acoustic flow function, which resulting in a large spatial smoothing neighborhood. Nonetheless, the motion acoustic flow of small target can assimilate due to it has no absolute dominant position compared with additional fast-moving ship, as shown in Figs~\ref{F7}(c)-(d). \\
\begin{table*}
\caption{\label{T3}AFBFs of the different spatial smoothing term.}
\begin{ruledtabular}
\begin{tabular}{lccccc}
&\multicolumn{5}{c}{Data set} \\
\cline{2-6}
Spatial smoothing term &Data 1 & Data 2 &Data 3 & Data 4 & Average \\
\hline
With non-local term    &0.0052 & 0.0053 &0.0070 & 0.0150 & 0.00813\\
\hline
w/o non-local term     &0.0118 & 0.0142 &0.0101 & 0.0232 & 0.01483\\
Contextual-based non-local term &$\mathbf{0.0050}$ & $\mathbf{0.0048}$ & $\mathbf{0.0065}$ & $\mathbf{0.0084}$ & $\mathbf{0.00618}$\\
\end{tabular}
\end{ruledtabular}
\end{table*}
\indent In order to analyze the improvement in MAFF estimation accuracy achieved by context-based non-local spatial smoothing term, AFBFs of MAFF estimation approach with different spatial smoothing term in motion acoustic flow function are further calculated. Motion acoustic flow function with non-local term is the basic configuration in this part (as \ref{E1}), which is defined as a baseline model 2. For each group of targets, we optimize the pre-processing method by the results of Sec.~\ref{subsubsec:4:2:1}. Based on Baseline Model 2, the spatial smoothing term in motion acoustic flow function is iteratively modified, including simply removing non-local term and contextual-based non-local term. As shown in Table \ref{T3}, approach with contextual-based non-local term in motion acoustic flow function achieves the lowest AFBFs among each approach. Comparing the AFBFs for each approach, it can be seen that as the motion estioation performance of Data set 4 achieves the biggest improvement, although it is also poorer than that of other Data sets.
\subsection{\label{subsec:4:3} Experiments results assessment: performance on small targets motion estimation}
\begin{figure*}
\begin{center}
\includegraphics[width = .9\textwidth]{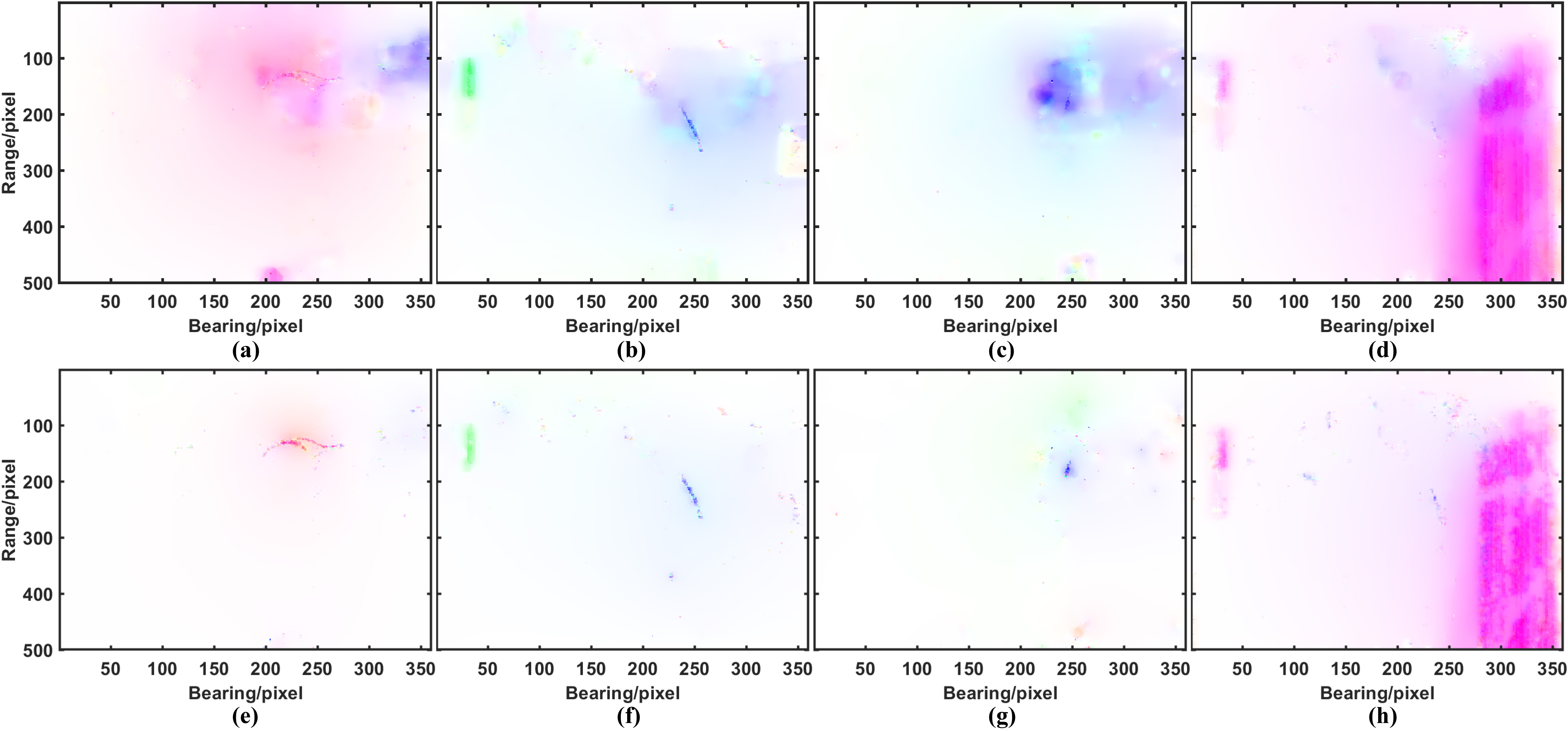}
\caption{\label{F8}{MAFF maps coded by different flow estimation approaches, (a) to (d) are the proposed MAFF estimation approach, (e) to (h) are the motion optical flow approach }}
\end{center}
\end{figure*}
\indent The results so far have indicated that proposed MAFF estimation approach has an improved accuracy than the motion optical flow approach. The following is the discussion on the application potential of the propose approach. The aim of this paper was to evaluate whether the modified motion acoustic flow function and calculate framework could effectively improve the estimated accuracy of the existing motion optical flow approach. In consideration of the practical needs of motion estimation, we explicitly show the MAFF maps (Normalized multi-frame cumulative MAFF) from the four data sets in Fig.~\ref{F7}. \\
\indent As shown in Fig.~\ref{F7}(e), the target 1 and 2 can be highlighted on a MAFF map in the form of a track, and more motion details are resumed compared with MAFF map estimated with motion optical flow approach in Fig.~\ref{F7}(a). Moreover, the number of pixels covered by moving clutter background is significantly reduced. Similarly, the proposed approach achieved higher accuracy in target 3 and 4, which move in straight lines in sonar echograph, as shown in Fig.~\ref{F7}(b)(c)(f) and (g). As shown in Fig.~\ref{F7}(d) and (h), the performance difference between the proposed approach and motion optical flow approach shows up after the intrude of fast-moving ship. The proposed approach preserves more motion feature of the target to avoid its MAFF being masked. \\
\indent The fine motion details are shown in MAFF maps real-time updated. The estimated initial range and speed information of the interested target in the detection area can be used as stated parameters for the extended Kalman filter (EKF) \citep{38}. On the other hand, object segmentation at object level need the motion feature as prior \citep{40}. With growing interest in using unsupervised learning for underwater target recognition, a high accuracy and robustness motion flow estimator may be used as input for subsequent processing.
\section{\label{sec:5} Conclusion}
\indent In this paper, a novel feature, MAFF, for characterizing the targets' motion mode in active sonar echograph of harbor environment are presented. The MAFF maps detect and preserve fine motion details of weak targets in real-world harbor environments in the South China Sea. In our experiment, the robustness of proposed method is tested by a series of targets in four data sets. In addition, the accuracy of MAFF estimation has been improved by new motion acoustic flow function with contextual-based non-local term and the calculation framework with pre-processing methods for sonar echograph.
\begin{acknowledgments}
\indent This work was supported in part by the National Key RandD Program of China under Grant No. 2016YFC1400200, in part by the National Natural Science Foundation of China under Grant No. 61671388, and in part by the National Engineering Laboratory for Integrated Aero Space-Ground-Ocean Big Data Application Technology.
\end{acknowledgments}





\bibliography{MAFF}

\end{document}